\documentclass[prl,twocolumn,showpacs,preprintnumbers,amsmath,amssymb,a4paper]{revtex4}
\usepackage{epsfig}
\usepackage{amsmath}
\usepackage{graphicx}

\begin{document}

\title{Macroscopic thermal entanglement due to radiation pressure}

\author{ Aires Ferreira$^{1,2}$  }
\author{ Ariel Guerreiro$^{1}$ }
\author{ Vlatko Vedral$^{2,3}$ }

\affiliation{ $^1$Faculdade Ci\^{e}ncias Universidade do Porto,
Rua do Campo Alegre 687 4169-007, Porto,
Portugal \\
$^2$Institut f\"{u}r Experimentalphysik, Universit\"{a}t Wien, Boltzmanngasse 5, A-1090 Vienna, Austria \\
$^3$The School of Physics and Astronomy, University of Leeds,
Leeds LS2 9JT, England }

\date{\today}

\begin{abstract}
Can entanglement and the quantum behavior in physical systems
survive at arbitrary high temperatures? In this Letter we show
that this is the case for a electromagnetic field mode in an
optical cavity with a movable mirror in a thermal state. We also
identify two different dynamical regimes of generation of
entanglement separated by a critical coupling strength.
\end{abstract}


\maketitle

Recently there has been a lot of interest in macroscopic systems
which can support non zero temperature entanglement
\cite{Vedral_b}\cite{Ghirardi}\cite{Vitali_Paris}. These works
suggest that entanglement not only persists in these conditions
but that it is also crucial in understanding the macroscopic
proprieties of the physical systems. On the other hand there is
also a possibility that, under some circumstances (very large
number of atoms, very large temperature, etc.), the physical
systems become effectively classical and we no longer require a
quantum description.

It is usually believed that decoherence explains the transition
from the quantum world to the classical world. Several other
alternatives, such as spontaneous wave-function collapse models
\cite{Ghosh} and gravitationally induced decoherence
\cite{Penrose} are also found in the literature. Different schemes
to create and probe macroscopical superpositions, which can give
us important clues about the quantum to classical transition, have
already been proposed for different physical
systems\cite{Mancini2}\cite{Bose}\cite{Others}\cite{Marshall}. For
instance, in \cite{Marshall}, like in the original
\textit{gedanken} experiment proposed by Schr\"{o}dinger, a single
photon state induces quantum superpositions of a mirror and in
\cite{Bose} multi-component cats of a cavity are created in the
interaction of a cavity field with a movable mirror.

Here we proceed in a different direction shedding light into the
problem of how quantum features may survive in the macroscopic
world. Recent research on this problem have shown that in systems
with finite Hilbert space dimension in equilibrium with a thermal
bath, bipartite entanglement vanishes above a critical temperature
\cite{Fine}. The following question arises, is this behavior
general, or are there some systems where entanglement is robust
against temperature?

In this Letter we will show that the entanglement between
macroscopic mirror and a cavity mode field can arise due to
radiation pressure at arbitrarily high temperatures as the system
evolves in time. This is very surprising because it is commonly
believed that high temperature completely destroys entanglement.
We will study entanglement in the time domain using a discrete
variable method and identify its dependence on the relevant
physical parameters, such as the strength of the radiation
pressure coupling and temperature.

In particular, we consider a perfect optical cavity with a movable
mirror with mass $m$ in one end, modeled as a mechanical harmonic
oscillator whose quivering motion is quantized. Identical systems
have been considered in the studying of decoherence \cite{Penrose}
and non classical states of the cavity field \cite{Bose}. The
general Hamiltonian of this system has been extensively studied by
Law \cite{Law}. Following the same approach as Law, we consider
the adiabatic limit where the resonant frequency of the mirror is
much slower than the frequency of the cavity mode $w_m<<2\pi n
c/L$, where $L$ is the length of the cavity when the mirror is in
equilibrium, $n$ is the order of the longitudinal cavity mode and
c is the speed of light. Henceforth, the coupling between
different cavity field modes (leading to the Casimir effect, etc.)
can be neglected. For cavities with very high quality factor Q the
damping is negligible, as it occurs on a time scale much longer
than it takes for the photons to perform several round trips.
Under these conditions the Hamiltonian includes only the free
terms of both the field and the mirror plus the interaction term
due to the radiation pressure (which causes the displacement of
the mirror),
\begin{eqnarray}
H=\hbar w_{0}a^{\dag}a+\hbar w_{m}b^{\dag}b-\hbar g
a^{\dag}a(b+b^{\dag}),
\end{eqnarray}
where $a$ is the annihilation operator of the cavity mode, $b$ is
the annihilation operator of the mirror, $w_0$ is the frequency of
the cavity mode and $g=w_{0}\sqrt{\hbar}/(L\sqrt{{m w_{m}}})$ is
the coupling constant.

The electromagnetic field is prepared in a coherent state,
$|\alpha\rangle$, using a driving laser tuned to resonance with
the cavity mode, whereas the mirror is initially in a Gibbs state
with temperature T. Then, the composite state of the system is
\begin{eqnarray*}
\rho(t_0)=\frac{1}{\mathcal{Z}}\int{\frac{d^2
z}{\pi}e^{-|z|^2/\bar{n}}
|\alpha\rangle\langle\alpha|\otimes|z\rangle\langle z|},
\end{eqnarray*}
where $\bar{n}=1/(e^{\hbar w_m/K_B T}-1)$ is the mean number of
excitations, $\mathcal{Z}$ is the mirror partition function and
$z$ represents all the possible coherent states of the mirror.

The evolution operator associated with the Hamiltonian (1) has a
closed formula and it was derived in \cite{Mancini2}, using the
Campbell-Baker-Hausdorff formula for the Lie algebra, and in
\cite{Bose}, using operator algebra methods:
\begin{eqnarray*}
U(t)=e^{-i w_0 a^{\dag} a t}e^{i k^2 (a^{\dag} a)^2 \Lambda(t)}
D_m(\eta(t) k a^{\dag} a) e^{-i w_m b^{\dag} b t},
\end{eqnarray*}
where $\Lambda(t)=w_mt - sin(w_m t)$, $\eta(t)=1-e^{-i w_m t}$,
$k=g/w_m$ and $D_m(\eta(t) k a^{\dag} a)=e^{k a^{\dag} a(\eta(t)
b^{\dag} -\eta(t)^* b)}$ is the displacement operator of the
mirror, $D_m(\gamma)|0\rangle=|\gamma\rangle$. Since the system is
periodic we only need to investigate entanglement in the time
interval $[0,2\pi/w_m]$.

The interaction term of the Hamiltonian has the potential to
entangle the cavity field modes with the vibrational modes of the
mirror. Heuristically, the generation of entanglement in this
physical system is better understood by considering the cavity in
the initial state $|0\rangle +|1\rangle$ and the mirror in the
vacuum state $|0\rangle_m$ (which is a good assumption for
$T\approx 0$). The state of the composite system evolves according
to (up to a normalization factor)
\begin{eqnarray*}
(|0\rangle + |1\rangle)\otimes|0\rangle_m \rightarrow |0\rangle
\otimes|0\rangle_m + e^{i f(t)}|1\rangle\otimes|k\eta(t)\rangle_m,
\end{eqnarray*}
where $f(t)$ is a phase, resulting in an entangled state for
$0<t<2\pi/w_m$. At $t=0$ and $=2\pi/w_m$, $\eta(t)$ becomes null
and the system returns to the initial separable state. The
entanglement results from the evolution of the term
$|1\rangle\otimes|0\rangle_m$, which can be interpreted as the
transference of momentum from the photon $|1\rangle$ to the mirror
$|0\rangle_m$, as the photon kicks the mirror. Though it is easy
to produce the state of light $|0\rangle +|1\rangle$
experimentally, the radiation pressure in this case is so small
that its virtually impossible to detect any entanglement using
present day technology. However, as we will show below, a
detectable amount of entanglement is expected when the cavity is
initially in a coherent state with sufficiently high amplitude.

Proceeding to the general case, we introduce the discrete variable
method which will allow the study of the entanglement in the
system. The density matrix $\rho(t)$ expressed in the Fock basis
of both cavity field and mirror reads:
\begin{eqnarray}
\rho = \sum_{\mu,\nu,n,m} \rho_{\mu \nu n m}|n\rangle\langle m|
\otimes |\mu\rangle\langle \nu|,
\end{eqnarray}
where the Latin indexes refer to the radiation and the Greek
indexes refer to the mirror. The elements of the density matrix
are calculated to be:
\begin{eqnarray}
\rho_{\mu\nu n m}=\left\{
\begin{array}
[c]{l}%
\mu! \delta_{\mu \nu} \Phi_{n m \mu \nu}(t)
e^{-\beta\hbar w_m(\mu+1)}\\
\quad\text{    , for $t=2 \pi k/w_m$ or $n=m=0$}\\ \\
(-1)^\mu \Phi_{n m \mu \nu}(t) e^{-\beta\hbar w_m(\mu+1)+\Omega_{n m}(t)} \\ W_{mn}^{*}(t)^{\nu-\mu} H_U[-\mu,1+\nu-\mu,z_{n m}(t)]\\
\quad\text{    , elsewhere.}%
\end{array}
\right.
\end{eqnarray}
$H_U[a,b,c]$ is the hypergeometric confluent function and
\begin{eqnarray*}
\Phi_{n m \mu \nu}(t)&=&\frac{\alpha^n\alpha^{\ast
m}e^{i\Lambda(t)(n^2-m^2)-|\alpha|^2}}{\mathcal{Z}(n!m!\mu!\nu!)^{1/2}}\\
W_{nm}(t)&=&\Delta_n(t)-e^{-\beta\hbar
w_m}(\Delta_n(t)+\Delta_m(t))/2\\
z_{nm}(t)&=&-e^{\beta\hbar w_m}W_{nm}(t)W_{mn}^{*}(t)\\
\Omega_{nm}(t)&=&(|\Delta_n(t)|^2+|\Delta_m(t)|^2)/2\\
&&\ \ -e^{-\beta\hbar w_m} |\Delta_n(t)+\Delta_m(t)|^2/4
\end{eqnarray*}
where $\Delta_n(t)=kn\eta(t)$.

Quantifying entanglement in mixed states is a non trivial problem,
except for bipartite two level systems, where the Peres criterium
is sufficient \cite{Peres} and necessary \cite{Horodecki} to
guarantee entanglement. In this Letter a new approach inspired by
\cite{Bose2} is followed. First, we project the original density
matrix (2) into a $2\times2$ subspace, which corresponds to a
local action ($P_{2\times2} \rho P_{2\times2}$) thus not
increasing the amount of entanglement $E(\rho)$ in the overall
system \cite{Vedral}, i.e. $E(\rho) \geq E(P_{2\times2} \rho
P_{2\times2})$. Afterwards different measures and markers of
entanglement for $2 \times 2$ systems can be applied. In
particular, the \emph{tangle} is an entanglement monotone valid
only for bipartite pure states \cite{Wootters}. The
\emph{negativity} is also an entanglement monotone but valid for
both pure and mixed bipartite states \cite{Vidal}
$\mathcal{N}(\rho)=(\sum_{i}{|\lambda_i|}-1)/2$, where
${\lambda_i}$ are the normalized eigenvalues of the partial
transposed matrix $\rho^{T_P}$. The relation between
\emph{negativity} and the Peres criterium is clear: it measures by
how much the partial transposed density matrix fails to be
positive, i.e. separable. Therefore to show the existence of
entanglement in the overall system it is sufficient to verify that
$E(P_{2\times2} \rho P_{2\times2})>0$. From now on we will use the
notation where each subspace of the density matrix is represented
by $[\mu,\nu;n,m]$ where $\mu,\nu$ refer to the number of
excitations of the mirror and $n,m$ refer to the number of
excitations of the cavity field.

For $T=0$ the system is in a pure state  and we can investigate
entanglement using the \emph{tangle}\cite{Wootters}, $\tau(t)$,
which is much simpler to compute than the \emph{negativity}.
Figure 1 shows that the system is always entangled except for
$t=0$ and $t=2\pi/w_m$ (only the range $t\in[0,\pi/w_m]$ has been
plotted since the \emph{tangle} has reflection symmetry around
$t=\pi/w_m$), when the mirror returns to its initial state. For
small $k$, the system reaches the maximum of entanglement at
$t=\pi/w_m$, simultaneously with the maximum displacement of the
mirror. For $k$ above a critical value, $k_c$, the maximum of
entanglement is achieved before $t=\pi/w_m$. The time of maximum
entanglement depends on the balance between the interaction time
$t_{int}$ ($t_{int} \sim 1/g$), i.e. the time scale of the
interaction term in the Hamiltonian, and the time of oscillation
of the mirror, $t_m$ ($t_m \sim 1/wm$).

It is interesting to try to understand the importance of the
amplitude of the coherent state, $\alpha$, in establishing the
value of $k_c$. We expect that when increasing $\alpha$ the value
of $k_c$ should decrease because there are more photons
interacting with the mirror resulting in a larger effective
coupling ($g \langle a^{\dag} a \rangle$). Surprisingly this is
not the case: the value of $k_c$ increases with $\alpha$. This can
be understood as follows. The ratio between the weight of the
$|n+1\rangle$ number state and the weight of $|n\rangle$ number
state in the expansion of the coherent state, being given by
$\alpha/(\sqrt{n+1})$, increases with $\alpha$, weakening the
entanglement generated after interaction with the mirror. The best
situation occurs when the weights of states are the most equally
distributed. Hence a higher coupling helps the entanglement
generation to have the same efficiency when $\alpha$ is increased.

Considering the subspace $[1,2;1,2]$, $k_c$ can be calculated as
solution of the transcendental equation
\begin{eqnarray}
-1+14k_c^2+24k_c^4=2\alpha^2(1+4k_c^2+96k_c^4)e^{-12k_c^2}\geq 0.
\end{eqnarray}
The right-hand side of equation (4) is non-negative resulting in a
restriction for $k_c$, i.e. $k_c$ is lower bounded. Also we can
see from equation (4) that $k_c$ increases with $\alpha$. This
confirms, at least for this subspace, that a higher coupling is
necessary for reaching the maximum of entanglement before
$t=\pi/w_m$ if the amplitude of the cavity field is increased.

Since $\tau(\pi)$ is the maximum of entanglement achieved during
the evolution of the system for $k\leq k_c$, and $k_c$ depends on
$\alpha$, the value of $\tau(\pi)$ is maximized for
\begin{eqnarray}
\alpha_{max}=\frac{e^{6k^2}\sqrt{1+2k^2}}{\sqrt{2}\sqrt{1+8k^2}},
\end{eqnarray}
which can only be defined for $k_c(\alpha_{max})\geq k$. Equation
(4) indicates that the ideal value of $\alpha$ increases with the
coupling. It is useful to define $\alpha_c$ as the value of
$\alpha$, for a given $k$, such that $k_c=k$ and to notice that
$k_c(\alpha_{max})\geq k$ is equivalent to the condition $\alpha_c
\geq \alpha_{max}$. By squaring (4) and dividing it by (5) (with
$k_c \rightarrow k$ and $\alpha \rightarrow \alpha_c$) the
condition of validity of (5) yields $k \geq 1/2$. This behavior is
independent of the particular subspace considered. For example,
the asymptotic behavior of the \emph{tangle} at $t=\pi$ as
function of $\alpha$ is always $\tau(\pi)\sim |\alpha|^2$, for
$\alpha \ll \alpha_{max}$, and $\tau(\pi)\sim |\alpha|^{-2}$, for
$\alpha \gg \alpha_{max}$.
\begin{figure}[h]
\centering
\includegraphics[scale=0.6]{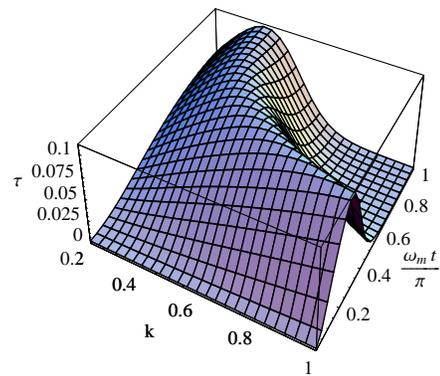}
\caption{\emph{Tangle} $\tau$ as a function of $k$ and the scaled
time for $T=0K, \alpha=1$ and subspace $[1,2;1,2]$.}
\end{figure}
Figure 1 shows how the maximum of entanglement in the considered
subspace is shifted to earlier times when $k$ increases.

For temperatures slightly above $T=0$ this behavior is not
significantly altered. At temperatures $T>0$ the system is in a
mixed state and the entanglement must be investigated using the
\emph{negativity}. It can be inferred from the plots of the
\emph{negativity} obtained from the simulations that, for the
subspace $[1,2;1,2]$, the value of $k_c$ increases slowly with the
temperature. The investigation of the \emph{negativity} also
shows, as expected, that the entanglement decreases with
temperature in a given subspace.

For higher temperatures, i.e. higher number of excitations of the
mirror, the \emph{negativity} cannot be computed because the
numerical calculation of the hypergeometric confluent function is
ill conditioned. However it is still possible to verify the
existence of entanglement for different values of $\alpha$, $k$
and temperature. This can be accomplished by introducing a marker
of entanglement based on the Peres criterium, which consists
simply in verifying whether the determinant of the partial
transposed density matrix $\rho^{T_P}$ is negative, thus
indicating the existence of a negative eigenvalue. This is of
course a very weak marker of entanglement, which can only detect
entanglement if $\rho^{T_P}$ has a odd number of negative
eigenvalues, but has the useful advantage of being easy to
calculate.

The determinant of $\rho^{T_P}$ for the subspace $[0,1;0,1]$ is $
\Upsilon_{[0,1;0,1]}(\alpha,b,x)=-G(\alpha,b,x) H(b,x)$, where
$H(b,x)=16x^2e^{b^2x}+x^2(-4+b^2(-2+x)(-1+e^{b^2/2}))^2-e^{b^2x/2}(b^4(-2+x)^4+32x^2+4b^2x^2(-1+e^{b^2/2})(4+(-3+e^{b^2/2})x))$,
$G(\alpha,b,x)=x^4|\alpha|^4e^{b^2(x-2)-4|\alpha|^2}/(16\overline{n}^4)$
is always positive, $b\equiv b(t)=\sqrt{2}k\sqrt{1-cos(w_m t)}$
and $x=\overline{n}/(1+\overline{n})$. The sign of $\Upsilon$ is
determined by $H(b,x)$ and is plotted as a function of
$(2/\pi)arctan(b(t))$ and $x$ in Figure 2, where it is clear that
the entanglement occurs for all parameters except for high values
of $b(t)$ and low temperature (region I) and for low values of
$b(t)$ and high temperatures (region II).

In the parameter region I, the lack of entanglement is not
important since, during the evolution of the system, the value of
$b(t)$ ranges between $0$ and $b_{max}=2k$ and, for sufficiently
short times, the system is entangled independently of how large
$k$ is. In the parameter region II the proof of existence of
entanglement in the system is not as easy because, in principle,
the system  could always be separable for small values of $k$.
However, by rewriting equation (3) as follows
\begin{eqnarray*}
\rho_{\mu \nu n m}=\Phi_{n m \mu \nu}(t)\int{\frac{d^2 z}{\pi}
F_n(z)^{\mu}F^{*}_m(z)^{\nu} e^{K_{n m}(z)}},
\end{eqnarray*}
where $K_{n m}(z)=-(|F_n(z)|^2+|F_m(z)|^2)/2-|z|^2/\overline{n}$
and $F_n(z)=z+k n \eta(t)$, it is clear that $k$ is always
multiplied either by $n$ or by $m$. Then it is straightforward to
verify that $\Upsilon_{[0,1;0,s]}$ it is proportional to
$\Upsilon_{[0,1;0,1]}$ if in the latter the coupling $k_s=k/s$ is
chosen, i.e.
\begin{eqnarray*}
|\alpha|^{-2s}s!
\Upsilon_{[0,1;0,s]}(k_s)=|\alpha|^{-2}\Upsilon_{[0,1;0,1]}(k).
\end{eqnarray*}
For high temperatures, if we choose a large $k$ capable of
producing entanglement in the subspace $[0,1;0,1]$, then there
must be entanglement in the subspace $[0,1;0,s]$ for the coupling
constant $k_s=k/s$, even though $k_s$ might not lead to
entanglement in the subspace $[0,1;0,1]$. Therefore, for systems
where $b(t)$ varies in region II there is always a $s$ such that
entanglement occurs in the subspace $[0,1;0,s]$.

\begin{figure}[h]
\centering
\includegraphics[scale=.35]{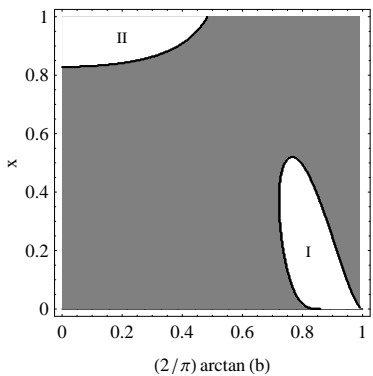}
\caption{Marker of entanglement, $\Upsilon$, as a function of
$(2/\pi)\arctan(b(t))$ and $x$ for the subspace $[0,1;0,1]$.
Entanglement (gray) exists for all parameter space except in
regions I and II (white).}
\end{figure}

Although entanglement occurs at any finite temperature this does
not imply the existence of entanglement in the limit of infinite
temperature. In fact, making use of the general inequality
$E(\sum_{i}{p_i \rho_{AB}^i}) \leq \sum_{i}{p_i E(\rho_{AB}^i)}$
to write $E(P\rho(t)P)\leq \int{dz e^{-|z|^2/\bar{n}} E(PU
|\alpha\rangle \langle\alpha|\otimes |z\rangle\langle z|
U^{\dag}P)/\overline{n}\pi}$ and taking into account that i) the
limit of infinite temperature is equivalent to the limit
$\overline{n} \rightarrow \infty$ and ii) entanglement is bounded
in any subspace, yields $\lim_{\overline{n}\rightarrow
\infty}E(P\rho(t)P)=0$.

The entanglement appears naturally in the dynamics of the system
for any value of coupling between the mirror and the cavity field
at any finite temperature, independently of the value of $\alpha$
(notice that $H$ does not depend on $\alpha$, at least for the
class of subspaces considered here), in spite of $\alpha$ playing
an important role in determining the amount of entanglement in the
system. An argument will be given to infer this role. We define
the normalized mutual information as $\mathcal{I}\equiv
I_{12}/(S_1+S_2)$, where $I_{12}\equiv S_1+S_2-S$, $S_i$ is the
Linear Entropy of the mirror ($i=1$) and the cavity field ($i=2$).
$S=1-1/(2\overline n+1)$ is the Linear Entropy of the composite
system and
\begin{eqnarray*} S_i=1-e^{-2|\alpha|^2}
f_i(\overline
n)\sum_{p,q=0}^{\infty}{\frac{|\alpha|^{2(p+q)}}{p!q!}e^{-g_i(\alpha,k,\overline
n,t)(p-q)^2}},
\end{eqnarray*}
for $t\neq 0,2\pi/w_m$, where $f_1(\overline
n)=1/(\overline{n}^2(a^2-1))$, $f_2(\overline n)=1$,
$g_1(\alpha,k,\overline n,t)=2k^2\xi(t)(1-b)$, $b=(1-1/a)^2+1/a$,
$a=1+1/\overline{n}$, $\xi(t)=1-\cos(w_m t)$ and
$g_2(\alpha,k,\overline n,t)=k^2\xi(t)(\overline n+2)$. Since
$g_i(\alpha,k,\overline n,t)>0$ (for $k\neq0$) we have $\partial
\mathcal{I}/
\partial |\alpha|>0$, which shows that quantum correlations
increase monotonously with $\alpha$ as mentioned before. Hence, a
detectable amount of entanglement is expected for sufficiently
high $\alpha$. Finally, we give a numerical value for $\mathcal I$
for realistic physical parameters $T=1K$,$w_m=10MHz$, $k=1$ and
$\alpha=10^6$, $\mathcal{I}-1/2\gtrsim0.19\times10^{-4}>0$. The
Araki-Lieb inequality \cite{Wherl} guarantees that quantum
correlations are present whenever $\mathcal{I}>1/2$. This is
consistent with our previous conclusions that entanglement
persists even in the high temperature limit.

In this Letter the generation of entanglement between a cavity
mode and a movable mirror via radiation pressure was analyzed by a
projection method. Since in the process of projection of the
density matrix much of the entanglement is lost, this method does
not allow to assess the entanglement of the overall system but it
is still robust enough to identify two dynamical regimes depending
on the coupling between the mirror and the cavity field. For
$k\leq k_c$ the maximum of entanglement is achieved at $t=\pi$ and
for $k>k_c$ is achieved before and furthermore demonstrate the
remarkable result that entanglement does occur for any finite
temperature.

We hope that our results stimulate further investigations on
entanglement in the macroscopic domain. We gratefully acknowledge
M. Aspelmeyer, \v{C}. Brukner, S. Gigan, J. Kofler, E. Lage, P.
Vieira and M. Wie\'{s}niak. A.F. is supported by FCT (Portugal)
through grant PRAXIS no. SFRH/BD/18292/04. V.V. acknowledges
funding from EPSRC and the European Union.

\end{document}